
\magnification=\magstep1
\overfullrule=0pt\nopagenumbers
\hsize=15.4truecm
\line{\hfil UQAM-PHE-94/03}
\centerline{\bf QCD CORRECTIONS TO THE $H^+\rightarrow
t\overline b$\ DECAY WITHIN THE}
\centerline{\bf MINIMAL SUPERSYMMETRIC STANDARD MODEL}
\vskip2cm
\centerline{HEINZ K\"ONIG\footnote*{email:konig@osiris.phy.uqam.ca}}
\centerline{D\'epartement de Physique}
\centerline{Universit\'e du Qu\'ebec \`a Montr\'eal}
\centerline{C.P. 8888, Succ. Centre Ville, Montr\'eal}
\centerline{Qu\'ebec, Canada H3C 3P8}
\vskip2cm
\centerline{\bf ABSTRACT}\vskip.2cm\indent
I present the results of the QCD corrections to the
$H^+\rightarrow t\overline b$\ decay within the
minimal supersymmetric standard model, if gluinos
and scalar quarks are taken within the relevant
loop diagram.
I include the mixing of the scalar partners of
the left- and right-handed top quark, which is
proportional to the top quark mass. The standard
corrections via gluons and quarks are about $+8$\%
for a charged Higgs mass of 300 GeV and about $-11$\%
for a Higgs mass of 800 GeV. I show that the standard
corrections are diminished or enhanced by a non-negligible
amount for certain values of the supersymmetric
parameter space. I also obtain sign changes.
\vskip1cm
\centerline{ May 1994}
\vfill\break
{\bf I. INTRODUCTION}\hfill\break\vskip.2cm\noindent
Recently there has been a lot of interest in the
QCD loop corrections [1,2]
as well as in the electroweek loop corrections [3,4]
of the charged Higgs boson decay into a top quark
and an anti-bottom quark.\hfill\break\indent
The underlying models were the two Higgs
doublet model [see
ref.5 and references therein] and the minimal supersymmetric
extension of the standard model (MSSM) [6,7].
In this Brief Report I take the last one as the underlying
model to consider the QCD corrections to the
$H^+\rightarrow t\overline b$\ decay mode.
\hfill\break\indent
According to eq.19 in ref.1 the QCD corrections via gluons
and quarks within the relevant loop diagram
(which I call SMG from now on) do have
both signs depending on the mass of the charged Higgs
boson. With the recently released CDF value for the
top quark mass [8] the ratio between the zeroth order
loop correction to the first order $\Gamma^1/\Gamma^0$\
is about $+7.5$\% for a charged Higgs mass of 300 GeV,
$-2.4$\% for a mass of 500 GeV and $-10.6$\% for a
mass of 800 GeV.\hfill\break\indent
In this Brief Report I present the QCD loop corrections to the
 $H^+\rightarrow t\overline b$\ decay if gluinos and scalar quarks
are taken within the loop as shown in Fig.1. Throughout
the calculations I neglect the mass of the bottom quark,
but I do not neglect the mixing of the scalar partners of
the left- and right-handed
top quark, which is proportional to the top quark
mass.\hfill\break\indent
In the next section I present the results of the calculation
of the Feynman diagram given in Fig.1 and in the last
section I discuss the results and end with the conclusions.
\hfill\break\vskip.2cm\noindent
 {\bf II. SUSY QCD CORRECTIONS TO $H^+\rightarrow t
\overline b$}\vskip.2cm
In the following I proceed as I did in the calculation
of the SUSY QCD corrections to  the $t\rightarrow H^+b$\
decay presented in ref.9. Taking the tree level coupling of
the charged Higgs to a top and bottom quark given in eq.(1)
in ref.9 the zeroth order to the decay rate
$\Gamma(H^+\rightarrow t\overline b)$\ is given by:
$$\Gamma^0(H^+\rightarrow t\overline b)=
{{G_F}\over{\sqrt{2}}}\vert
V^{KM}_{tb}\vert^2\cot^2\beta{1\over{4\pi}}
m_{H^+}\ m_{\rm top}^2\bigl (1-
{{m_{\rm top}^2}\over{m^2_{H^+}}}\bigr)^2\eqno(1)$$
$\cot\beta=v_1/v_2$\ is the ratio of the vacuum expectation
values (vev) of the two Higgs doublets and
 $V^{KM}_{tb}\approx 1$\
the value of the Kobayashi Maskawa matrix.
The calculation of the Feynman diagram in Fig.1 is similar
as presented in ref.9. As a result
I get for the first order in $\alpha_s$:
$$\eqalignno{\Gamma'^1(H^+\rightarrow t\overline b)=
&\Gamma^0
(H^+\rightarrow t\overline b)
\Bigl\lbrack 1-{{2\alpha_s}\over{3\pi}}
(S+A)\Bigr\rbrack&(2)\cr
S=&S_t+{{m_{\tilde g}}\over{m_{\rm top}}}S_{\tilde g}\cr
A=&A_t+{{m_{\tilde g}}\over{m_{\rm top}}}A_{\tilde g}\cr
S_t=&K_{11}\lbrack c^2_\Theta C_1^{\tilde b_1\tilde t_1}
+s^2_\Theta C_1^{\tilde b_1\tilde t_2}\rbrack
+K_{21}\lbrack s_\Theta c_\Theta(C_1^{\tilde b_1\tilde t_1}
-C_1^{\tilde b_1\tilde t_2})\rbrack\cr
A_t=&S_t\cr
S_{\tilde g}=&K_{11}\lbrack c_\Theta s_\Theta
(C_0^{\tilde b_1\tilde t_2}-
C_0^{\tilde b_1\tilde t_1})\rbrack-K_{21}\lbrack c_\Theta^2
C_0^{\tilde b_1\tilde t_2}
+s^2_\Theta C_0^{\tilde b_1\tilde t_1}\rbrack\cr
A_{\tilde g}=&S_{\tilde g}\cr
K_{11}=&1-{{m^2_W}\over{m^2_{\rm top}}}\tan\beta\sin 2 \beta \cr
K_{21}=&{1\over{ m_{\rm top}}}(A_{\rm top}+\mu\tan\beta)\cr}$$
$$\eqalignno{
C_0^{\tilde b_j\tilde t_i}=&-\int\limits_0^1d\alpha_1
\int\limits_0^{1-\alpha_1}d\alpha_2
{{m^2_{\rm top}}\over {f_{\tilde g}
^{\tilde b_j\tilde t_i}}}\cr
C_1^{\tilde b_j\tilde t_i}=&-\int\limits_0^1d\alpha_1
\int\limits_0^{1-\alpha_1}d\alpha_2
{{m^2_{\rm top}\alpha_1}\over{ f_{\tilde g}
^{\tilde b_j\tilde t_i}}}\cr
f_{\tilde g}^{\tilde b_j\tilde t_i}=&m^2_{\tilde g}-(m^2_{\tilde g}-
m_{\tilde t_i}^2)\alpha_1-(m^2_{\tilde g}-m^2_{\tilde b_j})\alpha_2-
m^2_{\rm top}\alpha_1(1-\alpha_1-\alpha_2)
-m^2_{H^+}\alpha_1\alpha_2\cr}
$$
where $c_\Theta=\cos\Theta$\ and $s_\Theta=\sin\Theta$ is
the mixing angle of the scalar top quarks with the
following mass matrix:
$$M^2_{\tilde t}=\left(\matrix{m^2_{\tilde t_L}
+m_{\rm top}^2+0.35D_Z^2&
-m_{\rm top}(A_{\rm top}+\mu\cot\beta)\cr-m_{\rm top}
(A_{\rm top}+\mu\cot\beta)&
m^2_{\tilde t_R}+m^2_{\rm top}+0.16D_Z^2\cr}\right)\eqno(3)$$
where $D_Z^2=m_Z^2\cos 2\beta$.
$m^2_{\tilde t_{L,R}}$\ are soft breaking masses,
 $A_{\rm top}$\ is the trilinear
scalar interaction parameter and $\mu$\ is the supersymmetric
mass mixing term of the Higgs bosons.
\hfill\break\indent
 The mass eigenstates
$\tilde t_1$\ and $\tilde t_2$\ are related to the current
eigenstates $\tilde t_L$\ and $\tilde t_R$ by
$\displaystyle{\tilde t_1=\cos\Theta\tilde t_L+\sin\Theta
\tilde t_R}$\ and $\displaystyle{\tilde t_2=\cos\Theta\tilde
t_L-\sin\Theta\tilde t_R}$. The mass eigenstates
are functions of the scalar breaking mass term $m_S$\ as well
as of $A_{\rm top}$\ and $\mu$. In a global SUSY model
we have $A_{\rm top}=m_S$\ and for neglecting bottom
quark masses we have
 $m^2_{\tilde b_1}=m^2_S-0.42m_Z^2\cos\ 2
\beta$\ and $m^2_{\tilde b_2}=m^2_S-0.08m_Z^2\cos\ 2\beta$.
With negelcting bottom quark mass
the scalar partners of the left and right handed bottom quarks
do not mix and therefore $m_{\tilde b_L}=
m_{\tilde b_1}$.
\hfill\break\indent
The S and A in eq.(2) indicate that the contribution
comes from the scalar and axial scalar coupling
of the matrix element. In the case of no mixing of
the scalar top quarks the gluino terms $S_{\tilde g}$\
and $A_{\tilde g}$\ do not contribute ($K_{21}=0=s_\Theta$).
The Feynman integration can be done numerically.
\hfill\break\indent
In eq.(19) in ref.1 the authors present the results of the
standard QCD one loop corrections,
which I have to include in my calculation.
As a final result I obtain:
$$\eqalignno{\Gamma^1(H^+\rightarrow t\overline b)=
&\Gamma^0(H^+\rightarrow t\overline b)
\Bigl\lbrack 1+{{2\alpha_s}\over{3\pi}} (G_{SM}-(S+A))
\Bigr\rbrack&(4)\cr
G_{SM}=&3{\rm ln}(1-\beta_t)-2{{(1-\beta^2_t)}\over{\beta_t}}
{\rm ln}{(1-\beta_t)}+2{\rm ln}(1-\beta_t)
{\rm ln}(\beta_t)\cr
&-(3+2\beta_t){\rm ln}(\beta_t)+{1\over 2}{\rm Li}_2((1-\beta_t)^2)
+3{\rm Li}_2(1-\beta_t)\cr &-4{\rm Li}_2((1-\beta_t)^{1/2})
-{2\over{\beta_t}}+{{13}\over 2}\cr
\beta_t=&1-{{m^2_{\rm top}}\over{m^2_{H^+}}}\cr}$$
In the next section I  discuss
the results.
\hfill\break\vskip.2cm\noindent
{\bf III. DISCUSSIONS}\vskip.2cm
To compare the standard QCD correction given in ref.1 with the
gluino and scalar contribution via eq.(2)
I present in Fig.2--5 the results of $\Gamma^1/\Gamma^0$\
for different cases of the charged Higgs mass, $\tan\beta$\
and the gluino mass as a function of the scalar mass $m_S$.
Throughout the calculation I use
174 GeV for the top quark mass and
$\mu=500$\ GeV with $A_{\rm top}=m_S$. In Fig.2 and Fig.3
I consider a charged Higgs mass of 300 GeV and in
Fig.4 and Fig.5 I take a charged Higgs mass of 800 GeV.
I use $\tan\beta=1$\ in Fig.2 and Fig.4 and $\tan\beta=2$\ in
Fig.3 and Fig.5. The results are presented for three different values
of the gluino mass 3 GeV (solid line), 100 GeV (dotted
line) and 500 GeV (dash-dotted line). The solid straight
line is the SMG contribution.\hfill\break\indent
One can see, that the gluino and scalar quark
contribution can change the SMG contribution drastically and even
lead to sign changes for certain values of the gluino
and scalar masses. In Fig.2 and Fig.4 the variation of the $\mu$\
hardly affects the results, whereas in Fig.3 and Fig.5 its
influence is much bigger due to a $\mu\tan\beta$\ dependence
in the couplings. For higher values of $\tan\beta$\ the
results are pushed farther away from the SMG result, although
in this case
as already was mentioned in ref.9, the bottom quark mass might
 become important. For $m_{H^+}=800$\ GeV the ratio
$\Gamma^1/\Gamma^0$\ is decreasing again if $m_S$\
is larger than 500 GeV.
\hfill\break\indent
For $\tan\beta=1$\ the heavier scalar top quark masses vary from
358 GeV to 631 GeV and the lighter scalar top quark mass
is about 250 GeV for $m_S$\ smaller than 100 GeV,
decreases constantly to 70 GeV for $m_S=350$\ GeV
and increases again to 260 GeV in the range considered here.
The scalar bottom masses are equal to $m_S$.
For $\tan\beta=2$\ the heavier scalar top quark masses
vary from 289 Gev to 594 GeV, the lighter scalar top
quark mass is about 145 GeV for $m_S$\ smaller than
100 GeV, decreases to 67 GeV for $m_S=250$\ GeV and
increases to 331 GeV for $m_S=450$\ GeV. The heavier
scalar bottom quark mass vary from 68 GeV to 452 GeV
and the lighter one from 54 GeV to 450 GeV in the
range of the scalar mass $m_S$\ considered here.
In both cases $\cos\Theta\approx 1/\sqrt{2}$.
\hfill\break\vskip.2cm\noindent
{\bf III. CONCLUSIONS}\vskip.2cm
In this Brief Report I have compared the first order
in $\alpha_s$\ contribution of the gluon and quarks
to the deacay rate $\Gamma(H^+\rightarrow t\overline b)$\
with the contribution of gluino and scalar quarks to
this decay rate. I have shown that the contribution
of scalar quarks and the gluino are not negligible
and changes the SMG contribution drastically and even
leads to different signs for certain values of the
SUSY masses.\hfill\break\indent
Finally I want to mention, that
the electroweak corrections with a top quark within the
loop can have a relative sign compared to the QCD corrections
depending on the charged Higgs mass and therefore also
might be important; this was shown in ref.3.
 Here in this Brief Report I have
shown that a gluino mass of 500 GeV contributes the most to
the $H^+\rightarrow t\overline b$\ decay rate and gets
even more important if $\tan\beta$\ increases whereas
according to ref.3 the electroweak top quark contribution
decreases.
\hfill\break\vskip.12cm\noindent
{\bf IV. ACKNOWLEDGMENTS}\vskip.12cm
I would like to thank the physics department
of Carleton university for the use of their computer
facilities. The figures were done with the very user
--friendly program PLOTDATA from TRIUMF.
\hfill\break\indent
This work was partially funded by funds from the N.S.E.R.C. of
Canada and les Fonds F.C.A.R. du Qu\'ebec.
\hfill\break\vskip.12cm\noindent
{\bf REFERENCES}\vskip.12cm
\item{[\ 1]}C.S. Li and R.J.Oakes, Phys.Rev.
{\bf D43}(1991)855.
\item{[\ 2]}A. M\'endez and A. Pomarol, Phys.Lett.
{\bf B252}(1990)461.
\item{[\ 3]}A. M\'endez and A. Pomarol, Phys.Lett.
{\bf B265}(1991)177.
\item{[\ 4]}J.M. Yang, C.S. Li and B.Q. Hu,
 Phys.Rev.{\bf D47}(1993)2872.
\item{[\ 5]}J.F. Gunion et al., "The Higgs Hunter's Guide"
(Addison-Wesley, Redwood City, CA, 1990).
\item{[\ 6]}H.P. Nilles, Phys.Rep.{\bf 110}(1984)1.
\item{[\ 7]}H.E. Haber and G.L. Kane, Phys.Rep.{\bf 117}(1985)75.
\item{[\ 8]}CDF Collaboration, Fermilab preprint, April 1994.
\item{[\ 9]}H. K\"onig, QCD corrections to the
$t\rightarrow H^+b$\ decay within the minimal supersymmetric
standard model, UQAM-PHE-94/01, hep-ph/9403297.
\hfill\break\vskip.12cm\noindent
{\bf FIGURE CAPTIONS}\vskip.12cm
\item{Fig.1}The diagram with scalar quarks and gluino
within the loop, which contribute to the charged Higgs boson decay
into a top and antibottom quark.
\item{Fig.2} The ratio of $\Gamma^1/\Gamma^0$\ as a function
of the scalar mass $m_S$\
for 3 different values
of the gluino mass: 3 GeV (solid line), 100 GeV (dotted line)
and 500 GeV (dash-dotted line) with $\mu=500$\ GeV and
$A_{\rm top}=m_S$. $v_1=v_2$\ and $m_{H^+}=300$\ GeV.
The solid straight line is the SMG contribution as given
in eq.(19) in [1].
\item{Fig.3}The same as Fig.2 with $v_2=2\cdot v_1$.
\item{Fig.4}The same as Fig.2 with $m_{H^+}=800$\ GeV.
\item{Fig.5}The same as Fig.4 with $v_2=2\cdot v_1$.
\vfill\break
\end
\end
\end